\newcommand{\Mpc}{h$^{-1}$ Mpc}
\newcommand{\rn}{$r_0$}

\documentstyle[12pt,aaspp4]{article}

\begin{document}

\title{Evolution of the Angular Correlation Function}
\author{A.J. Connolly, A.S. Szalay and R.J. Brunner}
\affil{Department of Physics and Astronomy, The Johns Hopkins
University, Baltimore, MD 21218}
\authoremail{ajc@skysrv.pha.jhu.edu,szalay@skysrv.pha.jhu.edu,rbrunner@skysrv.pha.jhu.edu}

\begin{abstract}
For faint photometric surveys our ability to quantify the clustering
of galaxies has depended on interpreting the angular correlation
function as a function of the limiting magnitude of the data. Due to
the broad redshift distribution of galaxies at faint magnitude limits
the correlation signal has been extremely difficult to detect and
interpret. We introduce a new technique for measuring the evolution of
clustering. We utilize photometric redshifts, derived from multicolor
surveys, to isolate redshift intervals and calculate the evolution of
the amplitude of the angular 2-pt correlation function. Applying these
techniques to the the Hubble Deep Field we find that the shape of the
correlation function, at $z=1$, is consistent with a power law with a
slope of $-0.8$. For $z>0.4$ the best fit to the data is given by a
model of clustering evolution with a comoving \rn = 2.37 \Mpc\ and
$\epsilon = -0.4^{+0.37}_{-0.65}$, consistent with published measures
of the clustering evolution. To match the canonical value of \rn = 5.4
\Mpc, found for the clustering of local galaxies, requires a value of
$\epsilon = 2.10^{+0.43}_{-0.64}$ (significantly more than linear
evolution). The log likelihood of this latter fit is 4.15 less than
that for the \rn = 2.37 \Mpc\ model. We, therefore, conclude that the
parameterization of the clustering evolution of $(1+z)^{-(3+\epsilon)}$ 
is not a particularly good fit to the data.
\end{abstract}


\section{Introduction}

The evolution of the clustering of galaxies as a function of redshift
provides a sensitive probe of the underlying cosmology and theories of
structure formation. In an ideal world we would measure the spatial
correlation function of galaxies as a function of redshift and type
and use this to compare with the predictions of different galaxy
formation theories.  Observationally, however, our ability to
efficiently measure galaxy spectra falls rapidly as a function of
limiting magnitude and consequently we are limited to deriving spatial
statistics from small galaxy samples and at relatively bright
magnitude limits (e.g.\ $I_{AB} < 22.5$, Le Fevre et al.\ 1996,
Carlberg et al.\ 1997).

To increase the size of the galaxy samples and thereby reduce the shot
noise the standard approach has been to measure the angular
correlation function, i.e.\ the projected spatial correlation function
(Brainerd et al.\ 1996, Woods and Fahlman 1997). While this allows us
to extend the measure of the clustering of galaxies to fainter
magnitude limits ($R<29$, Villumsen et al.\ 1997) it has an associated
limitation. For a given magnitude limit the amplitude of the angular
correlation function is sensitive to the width of the galaxy redshift
distribution, N(z). At faint magnitude limits N(z) is very broad and
consequently the clustering signal is diluted due to the large number
of randomly projected pairs.
 
In this letter we introduce a new approach for quantifying the
evolution of the angular correlation function; we apply photometric
redshifts (Connolly et al.\ 1995, Lanzetta et al.\ 1996, Gwyn and
Hartwick 1996, Sawicki et al.\ 1997) to isolate particular redshift
intervals. In so doing we can remove much of the foreground and
background contamination of galaxies and measure an amplified angular
clustering. We discuss here the particular application of this
technique to the Hubble Deep Field (HDF; Williams et al.\ 1996).

\section{The Photometric Catalog}

From version 2 of the ``drizzled'' HDF images (Fruchter and Hook 1996)
we construct a photometric catalog, in the $U_{300}$, $B_{450}$,
$V_{606}$ and $I_{814}$ photometric passbands, using the Sextractor
image detection and analysis package of Bertin and Arnout
(1996). Object detection was performed on the $I_{814}$ images using a
1 arcsec detection kernel. For those galaxies with $I_{814} < 27$ we
measure magnitudes in all four bands using a 2 arcsec diameter
aperture magnitude. The final catalog comprises 926 galaxies and
covers $\sim 5$ sq arcmin.

%We apply this
%conservative magnitude limit to ensure that the photometric
%uncertainties are small ($\Delta m < 0.1$) within each of the four
%passbands.
 
%Image detection was undertaken on each of the WFPC2 CCD chips
%independently. To derive a complete catalog of the full HDF frame we
%determine the astrometric solutions using the values given in Williams
%et al.\ (1996). We exclude overlapping regions between the WFPC2 chips
%due to the larger photometric uncertainties of galaxies detected
%within these areas and to avoid the problem of double counting
%galaxies. 

From these data we construct a photometric redshift catalog. For
$I_{814}<24$, we apply the techniques of Connolly et al. (1995,
1997), i.e.\ we calibrate the photometric redshifts using a training
set of galaxies with known redshift. For fainter magnitudes ($24<
I_{814} < 27$) we estimate the redshifts by fitting empirical spectral
energy distributions (Coleman et al.\ 1980) to the observed colors
(Gwyn and Hartwick 1996, Lanzetta et al.\ 1996 and Sawicki et al.\
1997). A comparison between the predicted and observed redshifts shows
the photometric redshift relation has an intrinsic dispersion of
$\sigma_z \sim 0.1$. 

\section{The Angular Correlation function}

We calculate the angular correlation function, $w(\theta)$, using the
estimator derived by Landy and Szalay (1993),
\begin{equation}
w(\theta) = \frac{DD - 2DR + RR}{RR},
\end{equation}
where $DD$ and $RR$ are the autocorrelation function of the data and
random points respectively and $DR$ is the cross-correlation between
the data and random points. In the limit of weak clustering this
statistic is the 2-point realization of a more general representation
of edge-corrected n-point correlation functions (Szapudi and Szalay
1997). As such it provides an optimal estimator for the HDF where the
small field-of-view makes the corrections for the survey geometry
significant.

We calculate $w(\theta)$ between 1 and 220 arcsec with logarithmic
binning. In the subsequent analysis we impose a lower limit of 3
arcsec to remove any artificial correlations due to the possibility
that the image analysis routines may decompose a single galaxy image
into multiple detections (at $z=1$ this corresponds to 12 h$^{-1}$ kpc
for $q_o = 0.5$).
%The upper limit represents the maximal separation
%between galaxy pairs that can be achieved within a single WFPC2 image.
For the random realizations we construct a catalog of 10000 points
(approximately 50 times the number of galaxies per redshift interval)
with the same geometry as the photometric data. 
%Increasing the number
%of random data points did not change the derived correlation function.
To account for the small angular size of the HDF we apply an integral
constraint assuming that the form of $w(\theta)$ is given by a power
law with a slope of $-0.8$.

Errors are estimated assuming Poisson statistics. The expected
uncertainty in each bin is calculated from the number of random pairs
(when scaled to the number of data points). Over the range of angles
for which we calculate the correlation function errors derived from
Poisson statistics are comparable to those from bootstrap resampling
(Villumsen et al.\ 1997).

%To account for the small angular size of the HDF we have to apply an
%integral constraint. This arises from the fact that we estimate the
%mean density from the observed number of galaxies, i.e.\ we have a
%conditional estimator given the number of galaxies. Thus, the average
%of $w(\theta)$ over whole area, estimated this way, is zero.  For
%narrow-angled pencil beam surveys this can lead to an underestimate of
%the clustering amplitude. The integral constraint, $C$, is given by,
%\begin{equation}
%C = \frac{1}{\Omega^2} \int \int d\Omega_1 d\Omega_2 w(\theta)
%\end{equation}
%where $\Omega$ is the solid angle that the HDF projects on the sky. For
%simplicity we assume that the form of $w(\theta)$ is given by a power
%law with a slope of $-0.8$ and fit $C$ for each redshift interval.

\subsection{The Angular Correlation Function in the HDF}

In Figure 1a we show the angular correlation function of the full
$I_{814}<27$ galaxy sample (filled triangles). The error bars
represent one sigma errors. The amplitude of the correlation function
is comparable to that found by Villumsen et al.\ (1997) for an R
selected galaxy sample in the HDF. It is consistent with a positive
detection of a correlation signal at the 2$\sigma$ significance
level. Superimposed on this figure is the correlation function for
those galaxies with $1.0<z<1.2$ (filled squares). Isolating this
particular redshift interval the amplitude of the correlation function
is amplified by approximately a factor of ten.

If we parameterize the angular correlation function as a power law
with $w(\theta) = A_w \theta^{1-\gamma}$ then, from Limber's equations
(Limber 1954), we can estimate how the amplitude, $A_w$, should scale
as a function of redshift and width of the redshift distribution,
\begin{equation}
A_w  = \sqrt{\pi} \frac{\Gamma[(\gamma - 1)/2]}{\Gamma[\gamma/2]} r_0^\gamma
\frac {\int_0^{\infty} dz N(z)^2 (1+z)^{-(3+\epsilon)} x(z)^{1-\gamma}  g(z)}
      {\left[ \int_0^\infty N(z) dz \right]^{-2}}
\end{equation}
where,
\begin{equation}
x(z) = 2\frac{((\Omega - 2) (\sqrt{1+\Omega z} -1) + \Omega z)}
{\Omega^2 (1+z)^2},
\end{equation}
is the comoving angular diameter distance, 
\begin{equation}
g(z) = (1+z)^2 \sqrt{1+\Omega z},
\end{equation}
N(z) is the redshift distribution and $\epsilon$ represents a
parameterization of the evolution of the spatial correlation function
(see below). 

For a normalized Gaussian redshift distribution, centered at
$\bar{z}$, with $\bar{z} \gg 0$ and dispersion $\sigma_z$, $A_w$ is
proportional to $1/\sigma_z$. Therefore, the amplitude of the angular
correlation function should be inversely proportional to the width of
the redshift distribution over which it is averaged.  In Figure 1 if
we assume that the magnitude limited sample ($I_{814}<27$) has a mean
redshift of $z=1.1$ and a dispersion of $\Delta z = 0.5$ (consistent
with the derived photometric redshift distribution) then isolating the
redshift interval $1.0<z<1.2$ should result in an amplification of a
factor of 5 in the correlation function (comparable to that which we
detect).

\subsection{The Angular correlation function as a function of redshift}

A limitation on studying large-scale clustering with the HDF is its
small field of view. For $\Omega=1$, the 220 arcsec maximal extent of
the WFPC2 images corresponds to 0.7 \Mpc\ and 0.9 \Mpc\ at redshifts
of $z=0.4$ and $z=1.0$ respectively.  Isolating very narrow intervals
in redshift (e.g.\ binning on scales of $\sigma_z <0.1$, the intrinsic
dispersion in the photometric-redshift relation) can, therefore,
result in the correlation function being dominated by a single
structure, e.g.\ a cluster of galaxies. To minimize the effect of the
inhomogeneous redshift distribution observed in the HDF (Cohen et
al. 1996) we divide the HDF sample into bins of width $\Delta z = 0.4$
based on their photometric redshifts.

For each redshift interval, $0.0<z<0.4$, $0.4<z<0.8$, $0.8<z<1.2$ and
$1.2<z<1.6$, we fit the observed correlation function, with a power
law with a slope of $-0.8$, over the range $3<\theta<220$ arcsec. From
this fit we measure the amplitude of the correlation function at a
fiducial scale of 10 arcsec. The choice of this particular angle is
simply a convenience as it is well sampled at all redshift
intervals. In Figure 2 we show the evolution of the amplitude as a
function of redshift. For redshifts $z>0.4$ the relation is relatively
flat with a mean value of 0.12. At $z<0.4$ we would expect the
amplitude to rise rapidly with redshift due to the angular diameter
distance relation. We find, however, that the amplitude remains flat
even for the lowest redshift bin. This implies that there is a bias in
the clustering signal inferred from the $0.0<z<0.4$ redshift interval
(see Section 4).

The value of the correlation function amplitude is comparable to those
derived from deep magnitude limited samples of galaxies. Hudon and
Lilly (1997) find an amplitude, measured at 1 degree, of $\log A_w =
-2.68 \pm 0.08$ for an $R<23.5$ galaxy sample. Woods and Fahlman for a
somewhat deeper survey, $R<24$, derive a value of $-2.94 \pm 0.06$. At
these magnitude limits the mean redshift is approximately 0.56 (Hudon
and Lilly, 1997) and the width is comparable to the redshift intervals
of $\Delta z = 0.4$ that we apply to the HDF data. Therefore, our
measured amplitude of $-2.92 \pm 0.06$ is in good agreement with these
previous results. 
%Extending the comparison to fainter magnitude limits
%(e.g. Brainerd et al.\ 1996) is not applicable due to the increasing
%width of the N(z) at faint magnitudes and the consequent dilution of
%the clustering signal.

\section{Modeling the Clustering Evolution}

We parameterize the redshift evolution of the spatial correlation
function as $(1+z)^{-(3+\epsilon)}$, where values of $\epsilon =
-1.2$, $\epsilon = 0.0$ and $\epsilon = 0.8$ correspond to a constant
clustering amplitude in comoving coordinates, constant clustering in
proper coordinates and linear growth of clustering respectively
(Peebles 1980). From Equation 2 we construct the expected evolution of
the amplitude of the angular correlation function, projected to 10
arcsec, for a range of values of $r_0$, $\epsilon$ and $\Omega$. We
assume that the intrinsic uncertainty of the photometric redshift for
each galaxy can be approximated by a Gaussian distribution, with a
dispersion $\sigma_z = 0.1$ (consistent with observations), and
determine the N(z) within a particular redshift interval as being
composed of a sum of these Gaussian distributions.
%corresponding to each galaxy in the survey.

In Figure 2 we illustrate the form of this evolution for two sets of
models, one with \rn =5.4 \Mpc\ and the second with \rn =2.37 \Mpc\
(the best fit to the data). For each model we assume $\Omega=1$ and
plot the evolutionary tracks for $\epsilon = -1.2$ (solid line),
$\epsilon = 0.0$ (dotted line) and $\epsilon = 0.8$ (dashed line). For
$z>0.4$ and a low \rn\ the observed amplitude of the correlation
function is well matched by that of the predicted evolution. For
redshifts $z<0.4$ the observed clustering is approximately a factor of
three below the \rn=2.37 \Mpc\ model.

This is not unexpected given the selection criteria for the HDF. The
field was chosen to avoid bright galaxies visible on a POSS II
photographic plate. This corresponds to a lower limit for the HDF
photometric sample of $F814W \sim 20$ (Marc Postman, private
communication). 
%The lack of bright galaxies suppresses the number of
%galaxy pairs at low redshift and thus the correlation function
%amplitude. From the Canada France Redshift Survey (CFRS, Lilly et al.\
%1995) we can estimate the redshift range over which we would expected
%the clustering signal to be biased. 
The redshift distribution for an $I_{814}<20$ magnitude limited sample
has a median value of $z=0.25$ and a width of approximately $\Delta z
= 0.25$ (Lilly et al.\ 1995). Therefore, by excluding the bright
galaxies within the HDF we artificially suppress the clustering
amplitude in the redshift range $0.0<z<0.5$. We can expect, as we have
found, that the first redshift bin in the HDF will significantly
underestimate the true clustering signal. Those redshifts bins at
$z>0.5$ are unlikely to be significantly affected by this magnitude
limit.

To constrain the models for the clustering evolution we, therefore,
exclude the lowest redshift point in our sample, i.e.\ $0.0<z<0.4$,
and determine the goodness-of-fit of each model using a $\chi^2$
statistic. The three dimensional $\chi^2$ distribution was derived for
the phase space given by $1<r_0<5$ \Mpc, $-4<\epsilon<4$ and
$0.2<\Omega<1$. We find that $\epsilon$ is relatively insensitive to
the value of $\Omega$ with a variation of typically 0.4 for the range
$0.2<\Omega<1.0$. As this is small when compared to the intrinsic
uncertainty in measuring $\epsilon$ we integrated the probability
distribution over all values of $\Omega$.

In Figure 3a we show the range of possible values for $\epsilon$ as a
function of \rn. The errorbars represent the 95\% confidence intervals
derived from the integrated probability distribution. Figure 3b shows
the log likelihood for these fits as a function of \rn.  The HDF data
are best fitted by a model with a comoving $r_0 = 2.37$ \Mpc\ and
$\epsilon = -0.4^{+0.37}_{-0.65}$. The value of \rn\ is comparable to
recent spectroscopic and photometric surveys with Hudon and Lilly
(1996) finding \rn = 2.75 $\pm 0.64$ \Mpc\ and Le F\'{e}vre et al.\
(1996) \rn = 2.03 $\pm 0.14$ \Mpc.

Given our redshift range, the I-band selected HDF data are comparable
to a sample of galaxies selected in the restframe U ($z=1.4$) through
V ($z=0.6$). To tie these observations into the clustering of local
galaxies we, therefore, compare our results with the B band selected
clustering analysis of Davis and Peebles (1983) and Loveday et al.\
(1992). Assuming a canonical value of \rn = 5.4 \Mpc\ we require
$\epsilon = 2.10^{+0.43}_{-0.64}$ to match the high redshift HDF data
(i.e.\ significantly more evolution than that predicted by linear
theory).

A bias may be introduced into the analysis of the clustering evolution
due to the fact that the I band magnitude selection corresponds to a
selection function that is redshift dependent (see above). If, as is
observed in the local Universe, the clustering length is dependent on
galaxy type then selecting different inherent populations may mimic
the observed clustering evolution. To determine the effect of this
bias we allow \rn\ to be a function of redshift (with \rn\ varying by
2 \Mpc\ from $z=0$ to $z=2$). The magnitude of this change in \rn\ is
consistent with the morphological dependence of \rn\ observed locally
(Loveday et al.\ 1995, Iovino et al.\ 1993). Allowing for this
redshift dependence reduces the value of $\epsilon$ by approximately
0.5 for all values of \rn\ (e.g.\ for \rn = 5.4 \Mpc\ $\epsilon =
1.6^{+0.43}_{-0.64}$).

It is worth noting that even with these large values of $\epsilon$ and
accounting for the bias due to the I band selection the evolution of
the clustering in the HDF is better fitted by a low value of \rn\ (the
log likelihood is 4.15 less than the fit to \rn =2.37
\Mpc). Parameterising the evolution of galaxy clustering is,
therefore, not particularly well represented by the form
$(1+z)^{-(3+\epsilon)}$ and it may be better for future studies to
discuss the evolution in terms of the amplitude at a particular
comoving scale rather than \rn\ and $\epsilon$.

\section{Conclusions}

Photometric redshifts provide a simple statistical means of directly
measuring the evolution of the clustering of galaxies. By isolating
narrow intervals in redshift space we can reduce the number of
randomly projected pairs and detect the clustering signal to high
redshift and faint magnitude limits. Applying these techniques to the
HDF we can characterize the evolution of the angular 2 pt correlation
function out to $z=1.6$. For redshifts $0.4<z<1.6$ we find that the
amplitude of the angular correlation function is best parameterized by
a comoving \rn=2.37 \Mpc\ and $\epsilon = -0.4^{+0.37}_{-0.65}$.  To
match, however, the canonical local value for the clustering length,
\rn=5.4 \Mpc, requires $\epsilon = 2.1^{+0.4}_{-0.6}$, significantly
more than simple linear growth.

It must be noted that while these results are in good agreement with
those from published photometric and spectroscopic surveys (Le
F\`{e}vre et al 1996, Hudon and Lilly 1996) there are two caveats that
should be considered before applying them to constrain models of
structure formation. The small angular extent of the HDF (at a
redshift, $z=1$, the field-of-view of the HDF is approximately 0.9
\Mpc) means that fluctuations on scales larger than we probe will
contribute to the variance of the measured clustering (Szapudi and
Colombi 1996). Secondly, the requirement that the HDF be positioned
such that it avoids bright galaxies ($I_{814}<20$) biases our
clustering statistics by artificially suppressing the number of low
redshift galaxies (a bias that will be present in most deep
photometric surveys). Therefore, the clustering evolution in the HDF
may not necessarily be representative of the general field
population. Given this, there is enormous potential for the
application of this technique to systematic wide angle multicolor
surveys, such as the Sloan Digital Sky Survey,

\acknowledgments
We would like to thank Marc Postman and Mark Dickinson for helpful
comments on the selection and interpretation of the Hubble Deep Field
data.  We acknowledge partial support from NASA grants AR-06394.01-95A
and AR-06337.11-94A (AJC) and an LTSA grant (ASZ).

\clearpage

\begin{figure}
\plotone{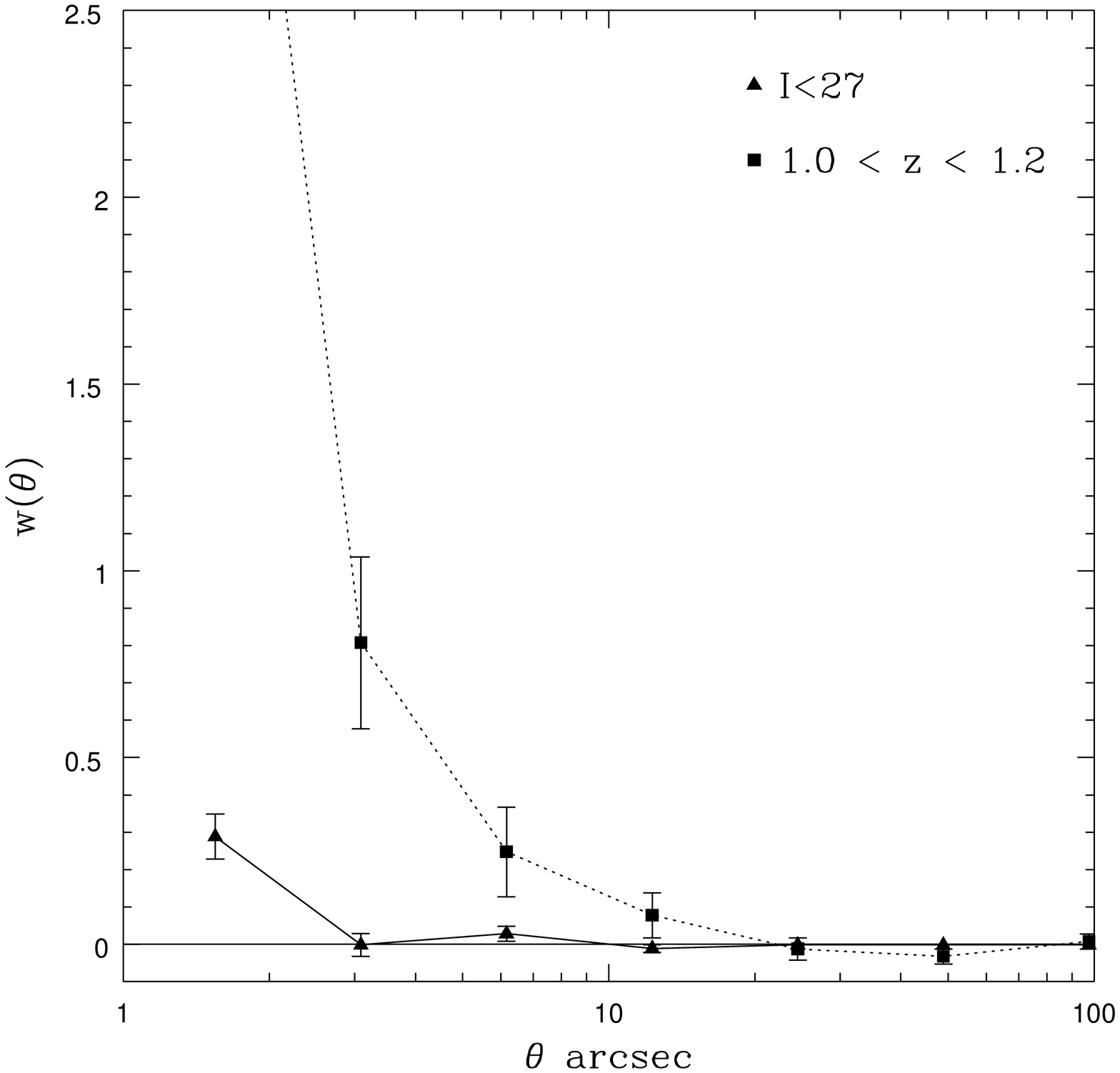}
\caption{The 2-pt angular correlation function for galaxies within the
HDF with $I_{814} < 27$. The triangles represent the correlation
function for a magnitude limited sample. The statistical significance
of a positive detection is approximately $2\sigma$. The squares show
the correlation function if we isolate, using photometric redshifts, a
subset of galaxies within the redshift interval $1.0<z<1.2$. The
amplification of the signal due to the reduction in the number of
projected random pairs is approximately a factor of 10.}
\end{figure}

\begin{figure}
\plotone{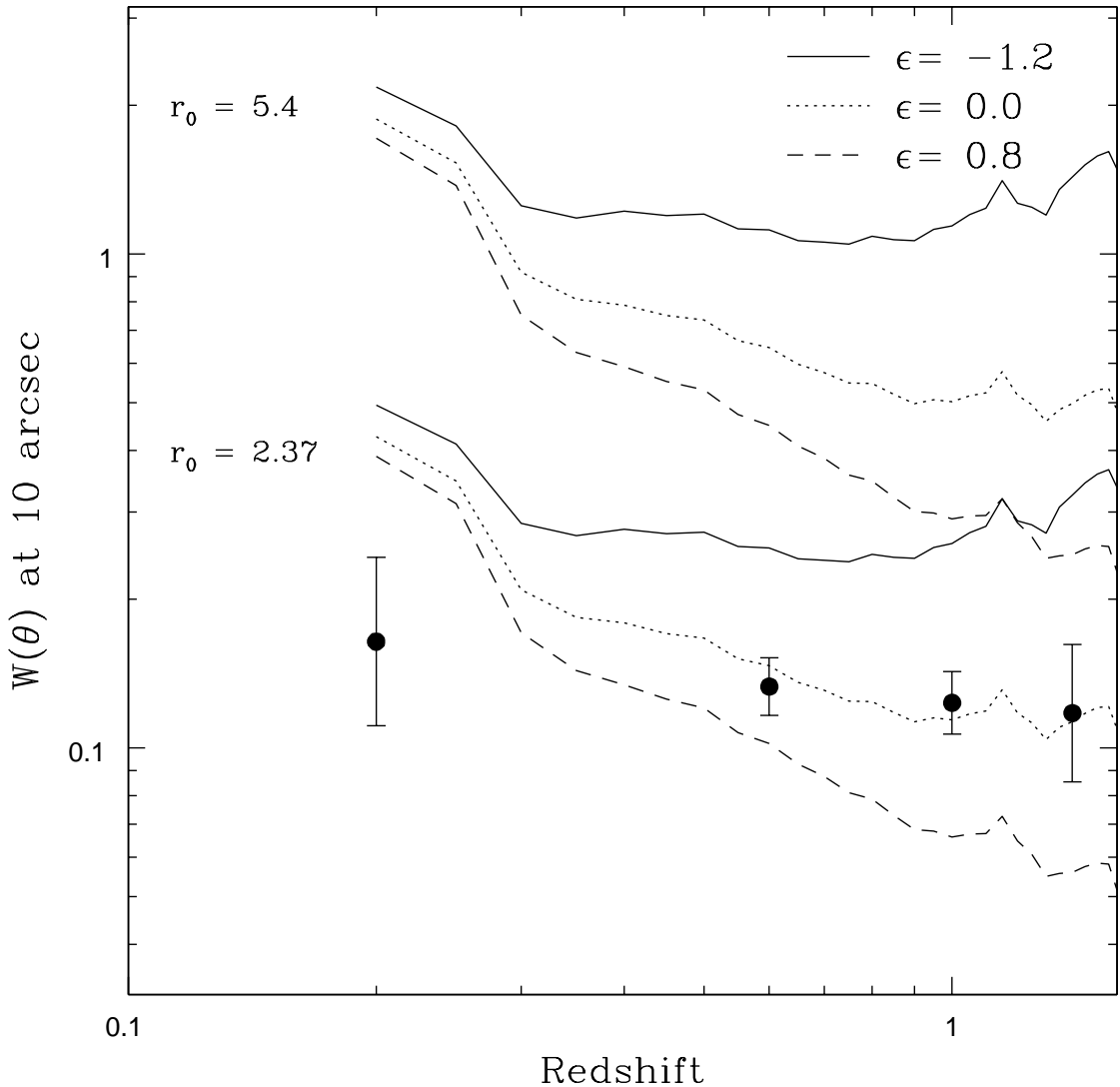}
\caption{Evolution of the amplitude of the angular correlation
function (measured at 10 arcsec) as a function of redshift.  Each
point is measured within a redshift interval of $\Delta z = 0.4$. The
exact N(z) for these intervals is constructed from the photometric
redshift distribution by assuming the error distribution for a
photometric redshift can be approximated by a Gaussian with a
dispersion of $\sigma_z =0.1$. We illustrate the expected evolution of
the correlation function amplitude for two sets of models, one with
$r_0 =2.37$ \Mpc\ (the best fit to the data) and a second with $r_0
=5.4$ \Mpc\ (the canonical value for local observations). For each
value of $r_0$ we give the evolution for $\epsilon=-1.2$ (fixed
clustering in comoving coordinates; solid line), $\epsilon=0.0$ (fixed
clustering in proper coordinates; dotted line) and $\epsilon=0.8$
(linear evolution of clustering; dashed line).}
\end{figure}

\begin{figure}
\plotone{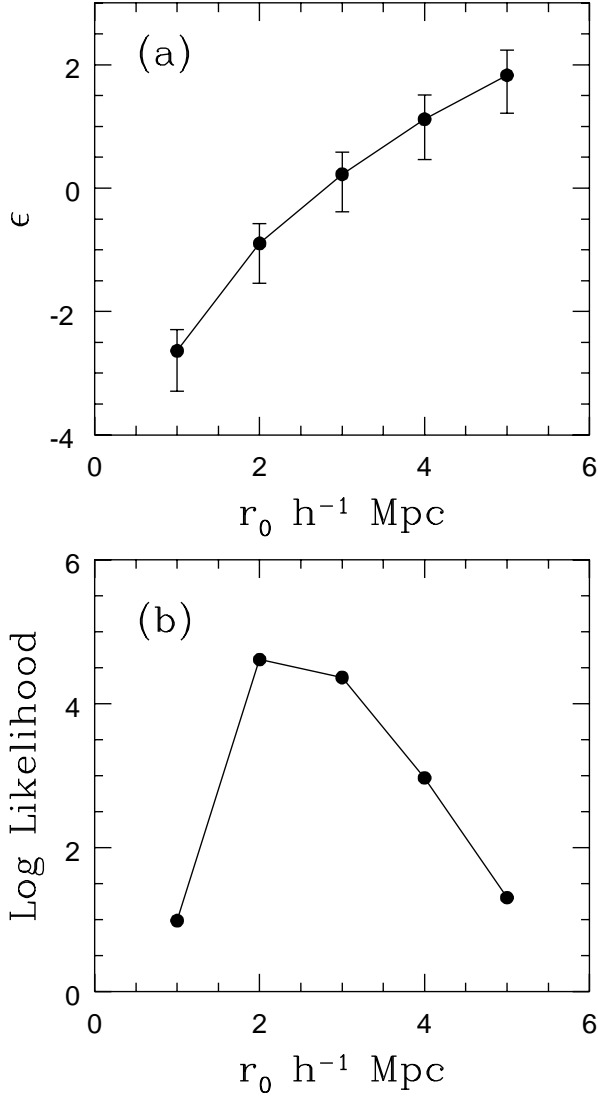}
\caption{(a) Integrating the probability distribution for the value of
$\epsilon$ at a given $r_0$ over all values of $0.2<\Omega<1.0$ we can
estimate the range of values of $\epsilon$ that best fit the HDF data.
For each value of $r_0$ within the range $1<r_0 <5$ \Mpc\ we give the
best fit for $\epsilon$ and the 95\% probability error bars. To fit
the local value of \rn = 5.4 \Mpc\ requires an $\epsilon =
2.1^{+0.4}_{-0.6}$, significantly more than linear evolution. (b) The
log likelihood for the best fit for $\epsilon$, at a given \rn. The
best fit to the data is given by an \rn = 2.37 \Mpc. The fit to \rn =
5.4 \Mpc\ has a log likelihood 4.15 less than that for \rn = 2.37
\Mpc.}
\end{figure}

\end{document}